# Open Medical Gesture: An Open-Source Experiment in Naturalistic Physical Interactions for Mixed and Virtual Reality Simulations

*Thomas B Talbot [1], Chinmay Chinara [1]*

*[1] University of Southern California Institute for Creative Technologies*

*Playa Vista, CA 90094, USA*

**ABSTRACT**

Mixed Reality (MR) and Virtual Reality (VR) simulations are hampered by requirements for hand controllers or attempts to perseverate in use of two-dimensional computer interface paradigms from the 1980s. From our efforts to produce more naturalistic interactions for combat medic training for the military, USC has developed an open-source toolkit that enables direct hand controlled responsive interactions that is sensor independent and can function with depth sensing cameras, webcams or sensory gloves. Natural approaches we have examined include the ability to manipulate virtual smart objects in a similar manner to how they are used in the real world. From this research and review of current literature, we have discerned several best approaches for hand-based human computer interactions which provide intuitive, responsive, useful, and low frustration experiences for VR users.

# INTRODUCTION

The lack of an effective and direct user interface for many medical simulation systems is a source of frustration for users and a barrier to adoption for creators (Seymour, 2008). The practice of medicine is a physical, human-contact centered activity that does not lend itself well to cursors and menus when attempting to conduct simulated interactions for purposes of patient examination or for procedural interventions (Barry et al., 2005). Intelligent interface design, user feedback and proper human factors diligence can produce a well-thought out, intuitive, easy to use and practical natural user interface that can function with various styles of medical simulations. An effective interface need not be tied to a specific technology approach or sensor, so it was logical to extend our work to multiple sensor systems to exercise and validate this approach. The starting point was to consider desirable features of a natural user interface (NI).

Any NI needs to be designed around the user experience (Johnson, 2010). A NI must be context aware, adapt to distance between the user and sensor. It should respond to the number and engagement of users. Interactions must be simple and easy to learn and master while avoiding the misinterpretation of user intent. Excellent interfaces will give constant feedback, so people always know what is happening and what to expect as an outcome of their actions. The strongest interfaces benefit from user testing, which is why we conduct user tests often and early in development (Proffitt & Lange, 2013).

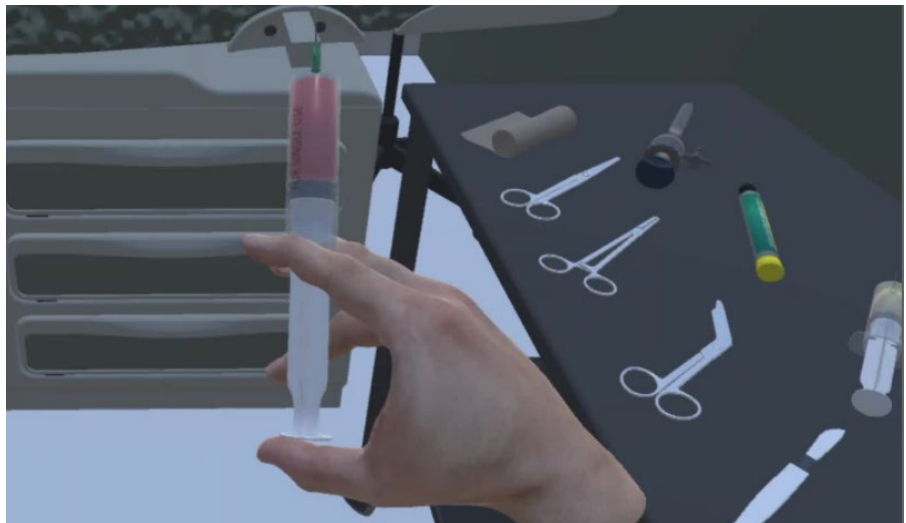
**Figure 1:** Gesture control of a syringe using a 3D camera.

# NATURAL INTERFACE STARTING POINTS

Effective interactions start with strong, unambiguous input methods that are precise, reliable, and fast while remaining considerate of sociological factors and understanding that users must feel comfortable using the input. Gestures themselves come in two varieties: innate and learned (Bannach et al., 2007). Innate gestures are things like pointing to aim, grabbing to pick up and pushing to select. These are facsimiles of real-world actions are already known to users. Learned gestures are things like 'press and hold to engage' and poses. Learned gestures are powerful but must be limited in number due to the learning curve associated with them. Gestures may be static, dynamic, or continuous actions.

When considering human factors with gestures, useful input from testers includes:

> *Did I learn all the basic gestures quickly?*
> *Now that I learned a gesture, can I quickly and accurately perform it?*
> *When I gesture, am I comfortable?*
> *When I gesture, is the application responsive?*
> *Does the application provide both immediate and ongoing feedback?*

From a Usability Engineering perspective four metrics are commonly used:

> **Learnability** – the time and effort for users to reach criterion performance
> **Throughput** – the speed of task execution and number of errors by experienced users
> **Flexibility** – the extent to which users can adapt a new system to new ways of interaction as they become more experienced
> **Attitude** – the degree to which a positive attitude is engendered in users by the system

Gestures must be designed for reliability. Programs should teach users how to effectively perform a gesture and instill confidence so that they can show others how to perform the gesture (Cabral et al., 2005). Users should be taught how to perform a gesture early in the experience so they can use it in similar contexts. Gestures must also be designed around natural interactions (Chaudhary et al., 2013); wrapping a bandage around a limb should be expressed with a wrapping motion, applying a tourniquet should be associated with a twisting or rotary motion, and cutting should be association with a linear motion.

A natural interface should allow people to interact from a distance. Gesture interfaces should ideally enable interactions and expressions that other input devices cannot, and they should fit the user's intended task. Finally, consideration should be given to adding voice to gestures as simple voice prompts can add several dimensions of flexibility to gestural interfaces.

# A FIRST ATTEMPT WITH MACHINE LEARNING

The first version of our gesture system tracked hand positions and employed a machine learning model to recognize static gestures and gesture in motion. It included a library of 46 gestures that were symbolic of medical tools and actions. For example, cupping the hand turned the hand into a stethoscope or a flat palm facing sideways with thumb abducting activates scissors.

As far as sensors went, a variety of depth sensing cameras and sensory gloves were tested. Good results were achieved with optical sensors such as the Leap Motion camera and Intel RealSense sensors. Microsoft Kinect yield poor hand interpretation performance, though newer versions of higher resolution would arrive in the future. For sensor gloves, there were few suitable for use, though tests were conducted with the Perception Neuron glove with a high degree of fidelity. Overall, optical sensors were better for determining hand location and orientation over long periods of time. They were excellent at determining gestures and finger placement though they were subject to poor identification when the hands are nearly perpendicular to the camera sensor or when obstructed by the other hand or an object. Glove sensors had superior interpretation of fingers & overall hand gesture, but they tended to require calibration at start of use and detection of absolute hand position became less accurate with prolonged use. Gloves had no issues with hand obstruction compared to cameras.

These learned gestures were captured quite accurately, though there was sometimes a delay in recognition of about one second or a requirement to perform some gestures more than once for reliable recognition. Symbolic recognition, hand position and orientation were tracked while accepting generalized input on the hand (open/closed) to actuate surgical forceps was also attempted (Figure 2).

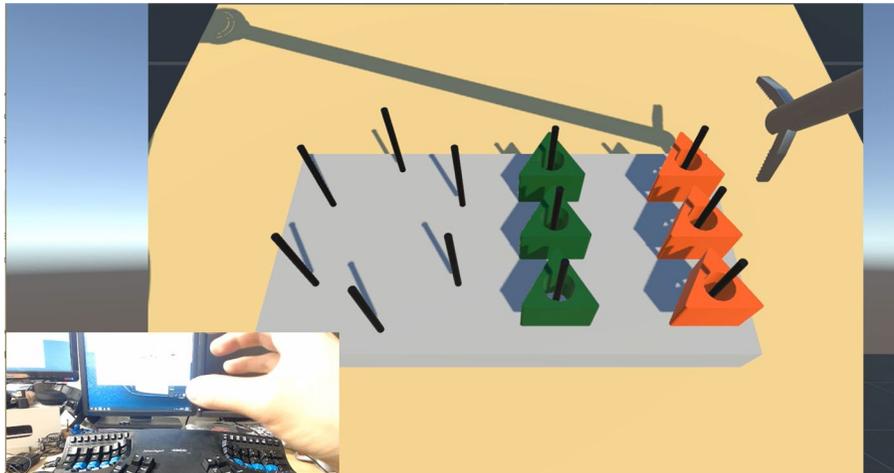

**Figure 2:** Fundamentals of Laparoscopic Surgery (FLS) task trainer simulated with symbolic gestures. The instrument follows hand position and orientation.

In OpenMG 1.0, the technology worked reasonably well but the requirement to memorize dozens of gestures was found to be prohibitive to adoption, so it was decided to pursue a different approach to gesture interpretation. The success of a symbolic recognition approach became the new starting point.

## OPENMG 2.0: REFINED METHODS

The center of an effective gesture system is a universal hand model that can map to inputs from several different kinds of sensors rather than depending on a specific commercial product. Parts of the hand are effectors in simulation space with a physics-based model. Therefore, translational and rotational forces from the hands will impact physical objects in VR which varies based on the mass of the virtual objects.

Computer code tied to virtual objects, called "Smart Objects", allowed the objects to have movement properties and collision detection for expected manipulation. Examples of smart objects include scissors, a ball, a turning knob, a moving lever, a syringe, or a human figure with moving limbs. Articulation points contain collision detectors and code to assist in expected hand actions. The OpenMG system includes a library of more than 40 Smart Objects in the toolkit. Thus, is it possible to throw a ball, hit that ball with a bat, cut a bandage, inject medications, turn on a ventilator or to lift and inspect a human arm.

It is important to mediate the interaction of the hands with virtual objects. Hands often violate the rules of a virtual world simply by passing through objects. One must interpret user intent. This can be achieved by introducing stickiness of the hands to objects. If the human's hands overshoot an object, we place the hand onto that object's surface unless the hand passes the object by a significant distance. Hands and fingers contact an object according to the object's contours and do not allow fingers to sink into the interior of an object.

### Physics Model Approach Demonstrated

OpenMG focused on four types of activities for demonstration and evaluation of the physics model approach and demonstrated them in an interactive playground. First, is the control of pushbuttons, lever switches and rotary dials. The second demonstration employs a tennis ball which can be picked up, juggled, or passed between hands. The third demo produced a baseball catch game with a virtual character to practice catching pitches and returning a throw. The fourth demonstration was patient palpation and actuating medical instruments such as scissors, forceps, syringes, bandages, and other medical tools. User testing feedback revealed difficulty with the medical instruments because user hands would not naturally conform to these complex shapes without a sense of touch to guide hand and finger positioning. The solution was to abstract the hand whereby picking up an instrument turns the user's hand into that instrument with an optimized hand placement upon the object. The instrument now simply tracks hand position and orientation. Actuation of a medical instrument (ex. Cutting) will open/close some scissors or push a syringe plunger. The interface tool will respond to any opening or closing of fingers or the hand to control this activity. This is an optimal balance of natural control and fidelity. This approach interprets user intent rather than directly applying finger positioning on medical tools. To return to a natural hand, the user can drop a medical instrument by placing his palm down and spreading all fingers open. Thus, a large variety of instruments are enabled with a universal gesture while also minimizing the user learning curve.

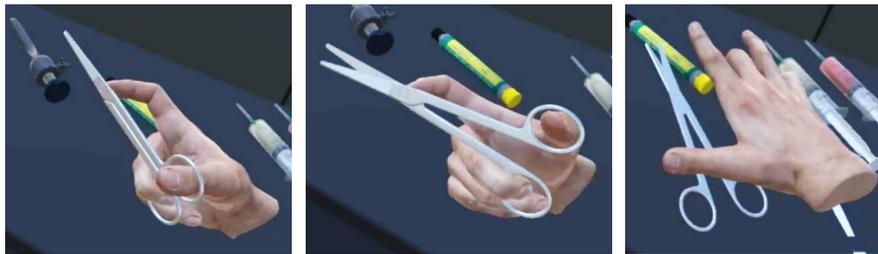

**Figure 3:** Scissor actuation by open and closing the hand. The system interprets intent & optimizes virtual hand position. Closed (left), Open (middle), Drop (right).

## Artificial Intelligence Techniques

The artificial intelligence (AI) technique to determine discrete hand gestures and motions within the physical space is a special form of AI called Long Short-Term Memory (LSTM). LSTM allows much faster and flexible recognition than other machine learning approaches. LSTM is particularly effective with points in motion. Latency of recognition is very low. In addition to LSTM, we employ other synthetic vision & object recognition AI to the discrimination of real-world objects. This allows for methods to conduct virtual simulations. For example, it is possible to pick up a virtual syringe and inject a medication into a virtual patient through hand motions while tracking hand points in contact with the virtual syringe. The system also detects when the hand is compressing the syringe plunger (Figure 1). One can also use virtual medications & instruments on human actors or manikins, not just on virtual objects. With object recognition AI, one can place a syringe on a tray in the physical world. The human user can pick up the syringe and use it on a virtual patient. Thus, anyone can blend physical and virtual simulation together seamlessly in a highly intuitive and naturalistic manner.

## Audio Synesthesia as a Haptic Substitute

Haptics, or a sense of physical resistance and tactile sensation from contacting physical objects is a supremely difficult technical challenge and is an expensive pursuit. One novel alternative approach ignores true haptics, called audio tactile synesthesia whereby the sensation of touch is substituted for that of sound. The idea is to associate parts of each hand with a tone of a specific frequency upon contacting objects. The attack rate of the sound envelope varies with the velocity of contact and hardness of the object being 'touched'. Such sounds can feel softer or harder (attack rate) depending on the nature of 'touch' being experienced. This substitution technique can provide tactile feedback through indirect, yet still naturalistic means.

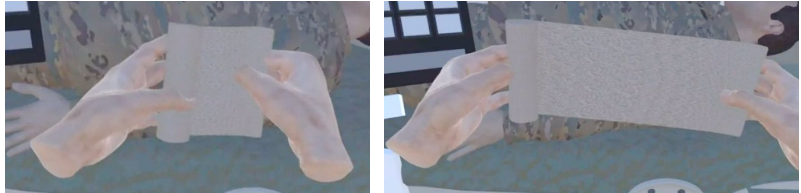

**Figure 4:** Hands stretch a bandage.

## OPTICAL SENSOR PLACEMENT

For desktop monitor applications, the optical sensor can be employed using a RealSense webcam integral to a laptop monitor or the sensor can be mounted vertically behind a keyboard. All approaches worked equally well. For VR/MR applications, we mounted the optical sensor on the front surface of a VR display or connected the sensor to a lanyard worn around the neck. The chest position of the lanyard provided superior sensor visibility for hands in most situations though the display mounted approach worked better for users who tended to turn their heads to extreme angles and contort themselves more. Experiments using HoloLens 2 employed onboard sensors within the wearable display, otherwise the Leap Motion sensor or Intel RealSense were utilized.

## CONCLUSION

The techniques and technologies explained here represent a baseline capability whereby interacting in mixed and virtual reality can now be much more natural and intuitive than it has ever been. Technology has now passed a threshold where game controllers and magnetic trackers are no longer necessary for VR. This advancement will contribute to greater adoption of VR solutions. Use of hands for direct actions makes use of VR more practical, easy and direct for users. To foster greater adoption, our team has committed to freely sharing Open Medical Gesture technologies for all purposes and at no cost as an open-source tool under the MIT Open-Source License. Such an approach permits academic or commercial use without royalties.

USC is in the process of integrating OMG technology into two commercial simulations (one surgical, one trauma) and one military medical simulation. The software toolkit will be made available to all at the AHFE 2022 conference.

In conclusion, we encourage the scientific, research, educational and medical communities to adopt these and other similar resources and determine their effectiveness and utilize these tools and practices to grow the body of useful VR applications.

## ACKNOWLEDGMENT

The authors would like to acknowledge the Joint Warfighter Medical Research Program for funding our research as well as the US Army Futures Command Soldier Training Technology Center, Orlando, FL and the office of Congressionally Directed Medical Research Program (CDMRP), Fort Detrick, MD for their outstanding guidance and cooperation. Initial funding from Army Research Laboratory Award W911NF-16-C-0033.